\documentclass[useAMS,usenatbib]{mn2e}
\usepackage{fancyhdr}
\usepackage{graphicx}
\usepackage{times}

\newcommand{\HI}{\hbox{\rmfamily H\,{\textsc i}}}
\newcommand{\HIsub}{\hbox{{\scriptsize H}\,{\tiny I}}}
\newcommand{\MHI}{\hbox{$M_{\HIsub}$}}
\newcommand{\rhoHI}{\hbox{$\rho_{\HIsub}$}}
\newcommand{\OHI}{\hbox{$\Omega_{\HIsub}$}}

\title[{\rm \HI} gas evolution at intermediate redshift]{Neutral atomic hydrogen ({\HI}) gas evolution in field galaxies at $z \sim$~ 0.1 and 0.2}
\author[J. Rhee et al.]{Jonghwan Rhee$^{1,6}$\thanks{E-mail: jrhee@mso.anu.edu.au}, Martin A. Zwaan$^{2}$, Frank H. Briggs$^{1,6}$,  Jayaram N. Chengalur$^{3}$, 
\newauthor Philip Lah$^{1,3,6}$, Tom Oosterloo$^{4,5}$, and Thijs van der Hulst$^{5}$ \\ 
$^{1}$Research School of Astronomy and Astrophysics, Australian National University, Canberra, ACT 2611, Australia\\
$^{2}$European Southern Observatory, Karl-Schwarzschild-Str. 2, Garching 85748, Germany\\
$^{3}$National Centre for Radio Astrophysics, Tata Institute for Fundamental Research, Pune 411 007, India\\
$^{4}$Netherlands Institute for Radio Astronomy (ASTRON), Postbus 2, 7990 AA Dwingeloo, The Netherlands\\
$^{5}$Kapteyn Astronomical Institute, University of Groningen, Landleven 12, 9747-AD Groningen, The Netherlands\\
$^{6}$ARC Centre of Excellence for All-sky Astrophysics (CAASTRO)}

\begin{document}

\date{Accepted 2013 August 06. Received 2013 August 06; in original form 2013 March 23}

\pagerange{\pageref{firstpage}--\pageref{lastpage}} \pubyear{2013}

\maketitle

\label{firstpage}

\begin{abstract}
We measure the neutral atomic hydrogen ({\HI}) gas content of field galaxies at intermediate redshifts of $z$~$\sim$~0.1 and $z$~$\sim$~0.2 using hydrogen 21-cm emission lines observed with the Westerbork Synthesis Radio Telescope (WSRT). In order to make high signal-to-noise ratio detections, an {\HI} signal stacking technique is applied: {\HI} emission spectra from multiple galaxies, optically selected by the CNOC2 redshift survey project, are co-added to measure the average {\HI} mass of galaxies in the two redshift bins. We calculate the cosmic {\HI} gas densities ({\OHI}) at the two redshift regimes and compare those with measurements at other redshifts to investigate the global evolution of the {\HI} gas density over cosmic time. From a total of 59 galaxies at $z$~$\sim$~0.1 we find {\OHI}~$=$~(0.33~$\pm$~0.05)$\times$10$^{-3}$, and at $z$~$\sim$~0.2 we find {\OHI}~$=$~(0.34~$\pm$~0.09)~$\times$~10$^{-3}$, based on 96 galaxies. These measurements help bridge the gap between high-$z$ damped Lyman-$\alpha$ observations and blind 21-cm surveys at $z=$~0. We find that our measurements of {\OHI} at $z$~$\sim$~0.1 and 0.2 are consistent with the {\HI} gas density at $z \sim$~0 and that all measurements of {\OHI} from 21-cm emission observations at $z \la$~ 0.2 are in agreement with no evolution of the {\HI} gas content in galaxies during the last 2.4~Gyr.
\end{abstract}


\section{INTRODUCTION}

Hydrogen gas in and around galaxies forms an important diagnostic for understanding galaxy formation and evolution. Since the Epoch of Reionisation (EoR) ended \citep[$z$~$\sim$~6,][]{Gunn:1965, Fan:2006}, the majority of the hydrogen gas in the intergalactic medium (IGM) and galactic halos has been highly ionised. Only a small fraction of the hydrogen resides in a neutral atomic phase when it is confined in gravitational potential wells, such as galaxies and damped Lyman-$\alpha$ absorption systems (DLAs). Since stars are formed from the neutral hydrogen gas trapped in galaxies, neutral hydrogen in galaxies is inextricably linked to the reservoir of star formation fuel as well as star formation.

While it is well established via many observations that over the last 8~Gyr (from $z$~$\sim$~1 to present), the rate of star formation in galaxies has declined by an order of magnitude \citep{Lilly:1996, Madau:1996, Hopkins:2004, Hopkins:2006}, little is known about the variation of neutral hydrogen gas, in particular during the period when stars formed actively.
Moreover, current observations of neutral hydrogen gas and star formation rate  in galaxies as a function of  redshift show that the amount of neutral hydrogen gas contained within galaxies is unlikely to be sufficient to produce the current stellar content and sustain on-going star formation \citep[e.g.,][]{Kennicutt:1998,Hopkins:2008,Bauermeister:2010,Leitner:2011}. 
To explain this discrepancy between the observed evolution of hydrogen gas and stellar contents in galaxies, a variety of ideas has been proposed and tested, such as gas accretion from the IGM and galactic halos, infalling gas associated with mergers and interactions, and galactic fountains to replenish star formation fuel \citep{Keres:2005,Sancisi:2008,Marinacci:2010}. 

A complicating factor is that observations to constrain the neutral hydrogen gas content in galaxies have been made using various approaches.
In the local universe, neutral hydrogen gas has been quantified using the {\HI} 21-cm emission line, corresponding to the hyperfine transition in the ground state of the hydrogen atom. Observations of {\HI} 21-cm emission have constrained the hydrogen gas content of nearby galaxies with fairly good precision \citep{Briggs:1990, Zwaan:1997, Zwaan:2003, Zwaan:2005, Martin:2010}. However, beyond the nearby Universe, it has been poorly constrained by observations of the 21-cm {\HI} emission line because of the inherent weakness  of the 21-cm emission line and the moderate sensitivity of radio astronomical instruments currently available. Hence, completely different observational strategies are used for high-redshift {\HI} measurements. Neutral hydrogen gas at high redshifts is primarily inventoried through the study of DLAs \citep[e.g.,][]{Wolfe:1986, Turnshek:1989, Lanzetta:1991}. These DLAs are the highest column density Lyman-$\alpha$ absorption features seen in the spectrum of quasars,
and from an unbiased DLA survey, one can determine the cosmic density, {\OHI}, of neutral hydrogen \citep{Wolfe:1986}. Since these early surveys, the number of DLA detections has risen to several thousands, leading to an accurate
measurement of {\OHI} for 1.5~$\la$~$z$~$\la$~5 \citep{Prochaska:2005, Prochaska:2009, Noterdaeme:2009, Noterdaeme:2012}.

Unfortunately, for various reasons, measuring neutral hydrogen gas with DLAs is less practical at redshifts ($z$~$\la$~1.5), corresponding to the time in cosmic history that showed a marked evolution of the star formation rate (SFR). In this redshift range,  Lyman-$\alpha$ absorption is detected at ultraviolet (UV) wavelengths, and it can thus be observed only from space. In addition, the incidence of DLAs per unit redshift is low below $z$~$\sim$~1.5, which makes it more difficult to statistically measure the hydrogen gas density. These difficulties result in the neutral hydrogen gas density measurements ({\OHI}) at intermediate redshifts being poorly constrained \citep{Rao:2006,Meiring:2011}. New or upgraded observing facilities operating in the radio wavelength regime may overcome these difficulties. While next generations of radio telescopes with high sensitivity are under development or in commissioning (e.g. JVLA\footnote{The Karl G. Jansky Very Large Array}, WSRT/APERTIF\footnote{APERture Tile In Focus}, ASKAP\footnote{Australian SKA Pathfinder}, MeerKAT\footnote{the Meer-Karoo Array Telescope}, SKA\footnote{Square Kilometre Array}), new approaches such as signal stacking and intensity mapping are endeavoured to measure hydrogen gas in galaxies at moderate redshifts and their feasibility has been proven \citep{Lah:2007, Lah:2009, Chang:2010,Masui:2013}.

Until the next generation radio telescopes comes online, the existing radio telescopes have been making the best use of their capabilities to measure {\HI} gas beyond the local Universe. For instance, since the first detection of {\HI} 21-cm emission from a cosmological distance using the Westerbork Synthesis Radio Telescope (WSRT) \citep[$z$~=~0.1766,][]{Zwaan:2001}, \citet{Verheijen:2007} observed two galaxy clusters, Abell~963 and Abell~2192, at $z$~$\sim$~0.2 using the WSRT with very long integration time to detect the {\HI} signal. As the Arecibo 305-m radio telescope's  front-end and back-end systems have been  upgraded recently, it gives access to frequencies down to 1120~MHz, corresponding to {\HI} redshifted to $z$~=~0.27. $\sim$20 isolated galaxies spanning the redshift range of $z$~=~0.17~--~0.25, pre-selected from the Sloan Digital Sky Survey (SDSS), were observed in order to measure {\HI} masses of individual galaxies \citep{Catinella:2008}. While these observing facilities push the limit of {\HI} detection into $z$~$\sim$~0.2, the Giant Metrewave Radio Telescope (GMRT) can reach {\HI} gas in galaxies up to $z$~$\sim$~0.4 by using signal stacking  \citep[$z$~=~0.24 and $z$~=~0.37,][]{Lah:2007,Lah:2009}, which is the same technical approach adopted in our research.

Previous {\HI} studies at intermediate redshifts have focused mostly on galaxy clusters \citep{Zwaan:2001, Verheijen:2007, Catinella:2008, Lah:2009}. Furthermore, on-going deep {\HI} surveys are pointing at galaxy clusters \citep[e.g. Blind Ultra Deep {\HI} Environmental Survey (BUDHIES),][]{Jaffe:2012} because the dense environment of a galaxy cluster is a good place to test galaxy evolution theories, for instance, the morphology-density relation \citep{Dressler:1980} and the Butcher-Oemler effect \citep{Butcher_Oemler:1978, Butcher_Oemler:1984}. However, concentrating on high density regions is likely to bias the investigation of the global trend of physical properties such as {\HI} gas density. {\HI} studies of normal environments, the so-called ``field'', are required to understand the variation of cosmological properties as well as the environmental effects on galaxy evolution. This current {\HI} study of field galaxies at $z$~$\sim$~0.1 and 0.2 hence both complements the {\HI} study of dense environments and will also help to improve our understanding of average properties of neutral hydrogen gas in the universe.

This paper is structured as follows. In Section~2 we detail the optical and radio data considered in this paper and also describe the radio data reduction. Section~3 describes individual detections and the stacking analysis to estimate {\HI} mass at $z$~$\sim$~0.1 and 0.2. The main results are presented in Section~4, and discussed in Section~5. Finally, the summary and conclusions are given in Section~6. We adopt the concordance cosmological parameters of $\Omega_{\Lambda}=$~0.7, $\Omega_{M}=$~0.3 and $H_{0}=$~70~km~s$^{-1}$~Mpc$^{-1}$  throughout this paper.

\begin{figure}
 \includegraphics[width=84mm]{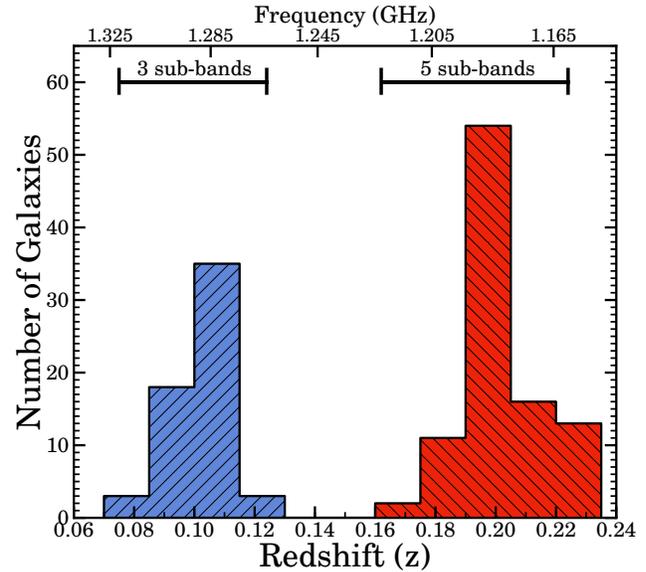}
 \caption{Histogram showing the redshift distribution of the 155 galaxies in the CNOC2~0920+37 catalogue that were observed with the WSRT. The red and blue indicate sub-samples at $z\sim$~0.1 and $z\sim$~0.2, respectively. The top x-axis shows frequencies the WSRT covered. The 3 higher frequency sub-bands of all sub-bands observed with the WSRT cover the sub-sample at $z\sim$~0.1 and the 5 remaining sub-bands with lower frequency include the sub-sample at $z\sim$~0.2.}
 \label{fig:hist}
\end{figure}
 
\section{THE DATA}
This research combines both optical and radio data, which are required for measuring the {\HI} signal using the signal stacking method. The {\HI} emission line spectra are taken from radio aperture synthesis observations and the complementary data, which provide spatial positions and redshifts, are obtained from optical photometric and spectroscopic surveys. 

\subsection{The optical data}

We selected one of the fields of the second Canadian Network for Observational Cosmology (CNOC2) Field Galaxy Redshifts Survey \citep{Yee:2000} which has been conducted using the 3.6~m Canada-France-Hawaii Telescope (CFHT) with the multi-object spectrograph (MOS). It provides a large number of redshifts and positions of field galaxies, spanning 0.0~$<$~$z$~$<$~0.6, as well as optical properties of galaxies, such as photometric measurements. More details regarding the observation and data reduction can be found in \citet{Yee:2000}. The CNOC2~0920+37 field selected for radio observation is one of four patches the survey covered, and the catalogue provides positions, redshift measurements and five-band photometry $U B V R_{c} I_{c}$ for 1630 galaxies. Of these, only 155 galaxies lie within the WSRT frequency and beam coverage. The redshift distribution of these galaxies is shown in Fig.~\ref{fig:hist}. We divide all galaxies used for analysis into two sub-samples (at $z$~$\sim$~0.1 and $z$~$\sim$~0.2) as shown in the redshift distribution plot.

\begin{figure}
 \centering
 \includegraphics[trim=0mm 0mm 0mm 0mm, clip, width=76mm]{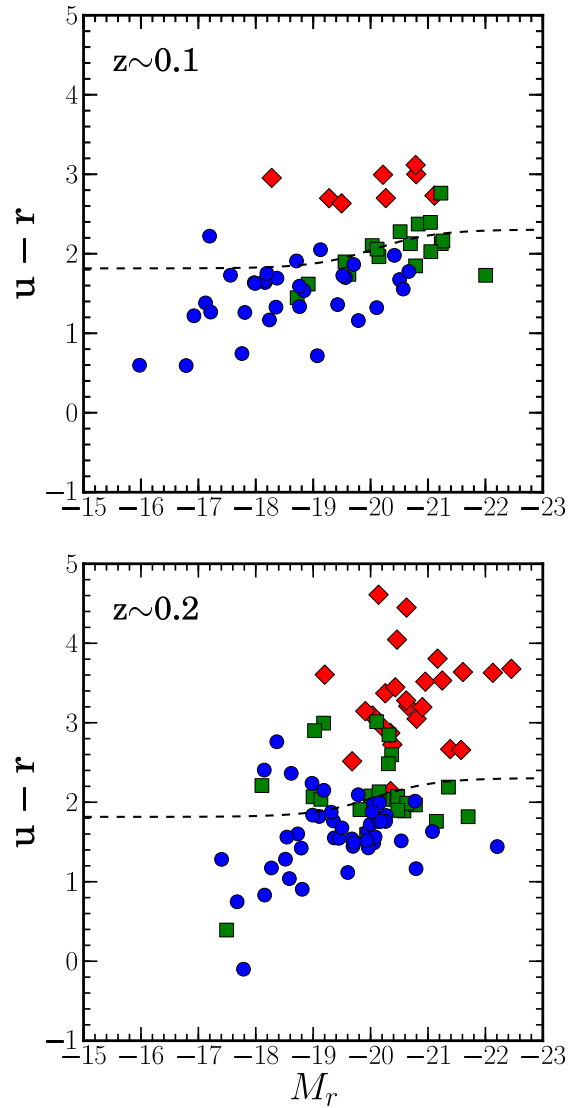}
 \caption{The colour-magnitude diagrams of the sample galaxies at $z\sim$~0.1 ({\it top}) and $z$~$\sim$~0.2 ({\it bottom}). The blue filled circles in each plot represent galaxies classified as late-type, the green filled squares are intermediate-type galaxies and the red filled diamonds are classified as early-type galaxies in the CNOC2~0920+37 catalogue using SED fitting of optical colours. The dashed line is the divider between the blue cloud and red sequence as determined by \citet{Baldry:2004}. Here, $M_{r}$ is the Petrosian absolute r-band magnitude and $u-r$ is the colour derived from the $u$-band and $r$-band model magnitudes from the SDSS.}
 \label{fig:cmd}
\end{figure}

The CNOC2~0920+37 catalogue provides only relative positions of the galaxies with respect to the reference centre of the CNOC2~0920+37 field instead of the actual right ascension (R.A.) and declination (Dec.). Since accurate galaxy positions are crucial for the stacking technique, we calculate the R.A. and Dec. from the relative positions in the CNOC2~0920+37 catalogue and then cross-match them against  the positions given in the Sloan Digital Sky Survey Data Release~7 \citep[SDSS DR7,][]{Abazajian:2009}. We compare coordinates from both catalogues and look at images of individual galaxies taken from the SDSS to check whether there is a mismatch. All 155 galaxies in our sample have positions that match within 8 arcsec with the SDSS DR7
positions. This positional error is small compared to the WSRT beam size which is about 4 times larger. We use the SDSS positions for extracting the spectra used in the stacking analysis. The SDSS catalog also provides us with $ugriz$ photometric magnitudes and measurements of the galaxy sizes (i.e. the Petrosian radius). The SDSS photometry of all galaxies used for the analysis in this paper is corrected for the effect of dust extinction from the Milky Way using the dust maps of \citet{Schlegel:1998}. The $k$-correction \citep{Blanton:2003,Blanton:2007} is then applied to convert all observed magnitudes to rest-frame magnitudes.     

The galaxies in the CNOC2~0920+37 catalogue are classified following two schemes. The first is a spectral classification, where the ``spectral class'' is based on spectroscopic inspection during the cross-correlation procedure for redshift estimation. The galaxy spectra are classified in four classes -- absorption line/early type, intermediate-late type, emission line and active galactic nuclei (AGNs). Our sample includes only the first three classes, and there is no object classified as an AGN. However, during cross-matching the CNOC2~0920+37 catalogue with the SDSS, we found that one of our sample galaxies is classified as a quasar in the SDSS although it belongs to intermediate-type in the CNOC2~0920+37 catalogue. The SDSS spectrum for this object shows characteristics of Seyfert galaxies, which are considerably less luminous than the typical quasar. We include it in our sample as an intermediate class galaxy for {\HI} density analysis. 

The second classification comes from fitting galaxy SEDs to the observed  colours. 
This classification scheme is based on least square fits of the SEDs derived via the methodology of \citet{Coleman:1980} to the CNOC2 colours. Based on these fits the sample galaxies are assigned to one of three spectro-photometric classes, viz. ``early'', ``intermediate'' and ``late''. In Fig.~\ref{fig:cmd} we compare this classification with the photometric classification done using SDSS photometry \citep{Baldry:2004}. As can be seen from the figure, the CNOC2 galaxy classification scheme seems to be in good agreement with the galaxy classification based on the SDSS photometry.  Most early-type galaxies in the CNOC2 catalogue are located in the red sequence, representing early-type galaxies on the SDSS colour-magnitude diagram (CMD). Those categorised as late-type galaxies in the CNOC2 are found in the area corresponding to the blue cloud area on the CMD. The intermediate galaxies are distributed between the red sequence and the blue cloud, which is similar with that referred to as green valley \citep{Wyder:2007}. We adopt the CNOC2 galaxy classification scheme based on SED fitting for our further analysis.

\begin{figure*}
 \includegraphics[trim=0mm 0mm 0mm 5mm, clip, width=130mm]{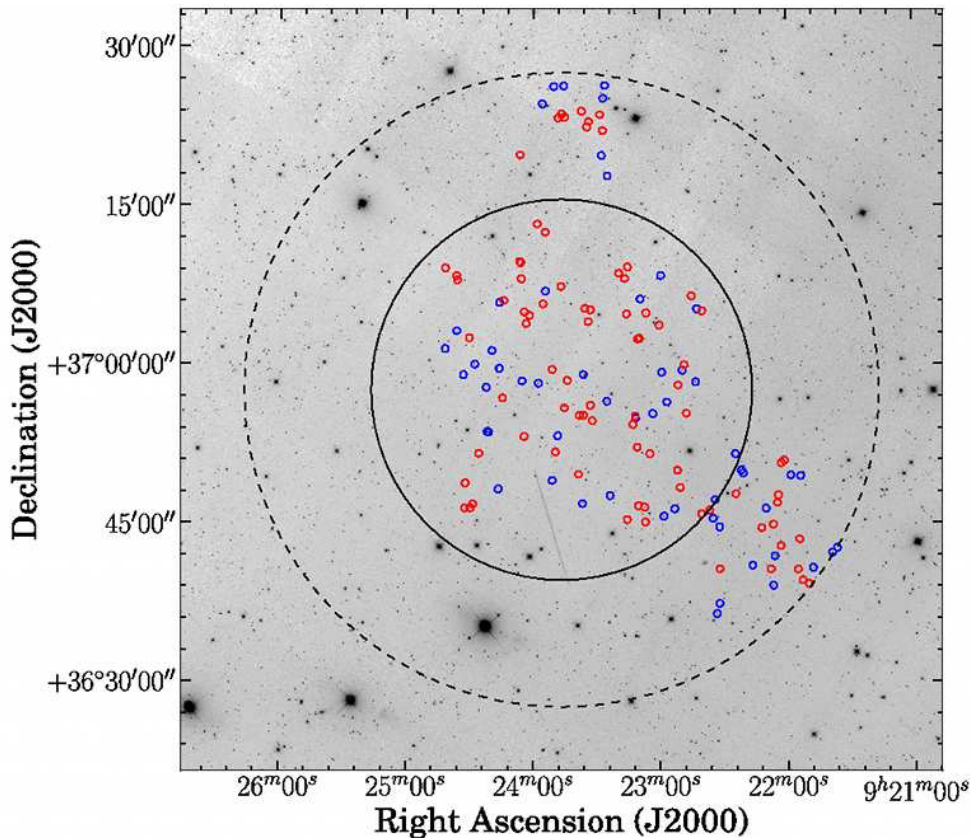}
 \caption{The pointing centre and the primary beam size of the WSRT are overlaid on a SDSS image of CNOC2~0920+37. The solid line and the broken line indicate the HPBW of the primary beam ($\sim$36~$\arcmin$) and 10 per cent level of the primary beam ($\sim$1.0$\degr$ in diameter), respectively. The blue and red circles indicate our sample galaxies located at $z$~$\sim$~0.1 and at $z$~$\sim$~0.2. }
 \label{fig:wsrtbeam}
\end{figure*}

\subsection{The radio data}

Radio observations of the CNOC2~0920+37 field were performed for 10~$\times$~12~hr (10~days) between 10 April and 12 May in 2003 using the WSRT. One 12~hr observation was excluded from the current analysis because its frequency configuration is inconsistent with the other days. This leaves 106.4 hours of telescope time to be used for on-source integration. The pointing centre of the WSRT observation for the CNOC2~0920+37 field is R.A.~09$^{\rmn h}$ 23$^{\rmn m}$ 46$\rmn \fs$507~Dec.~+36$^{\rmn d}$ 57$^{\rmn m}$ 37$\rmn \fs$432 (J2000). The WSRT primary beam covers the central area of the CNOC2~0920+37 as shown in Fig.~\ref{fig:wsrtbeam}. The standard correlator configuration gave partially overlapping 8~$\times$~20~MHz bands with 128 channels per band and 2 polarisations per channel. The channel spacing within each subband is 0.15625~MHz (37.9~km~s$^{-1}$ at $z=$~0.15). Hanning smoothing was applied to the spectral line data to suppress the Gibbs ringing phenomenon near the band edge and surrounding strong line signals. The observed frequency ranges span 1321~--~1160~MHz, which is a redshift range of 0.075~$<z<$~0.224 for redshifted {\HI} 21-cm emission. However, as seen in Fig.~\ref{fig:hist}, there is a gap within the frequency coverage to avoid the well-known severe radio frequency interference (RFI) zone, such as radar, GPS and radio amateur bands, affecting frequencies from 1.263~GHz to 1.222~GHz, corresponding to the redshift between 0.124 and 0.162. As described in more detail below, we divide our sample in two using a cut in redshift space, with galaxies with redshifts corresponding to the first three sub-bands in one sub-sample, and the galaxies with redshifts in the remaining five sub-bands in the second sub-sample.

\begin{figure*}
 \includegraphics[width=130mm]{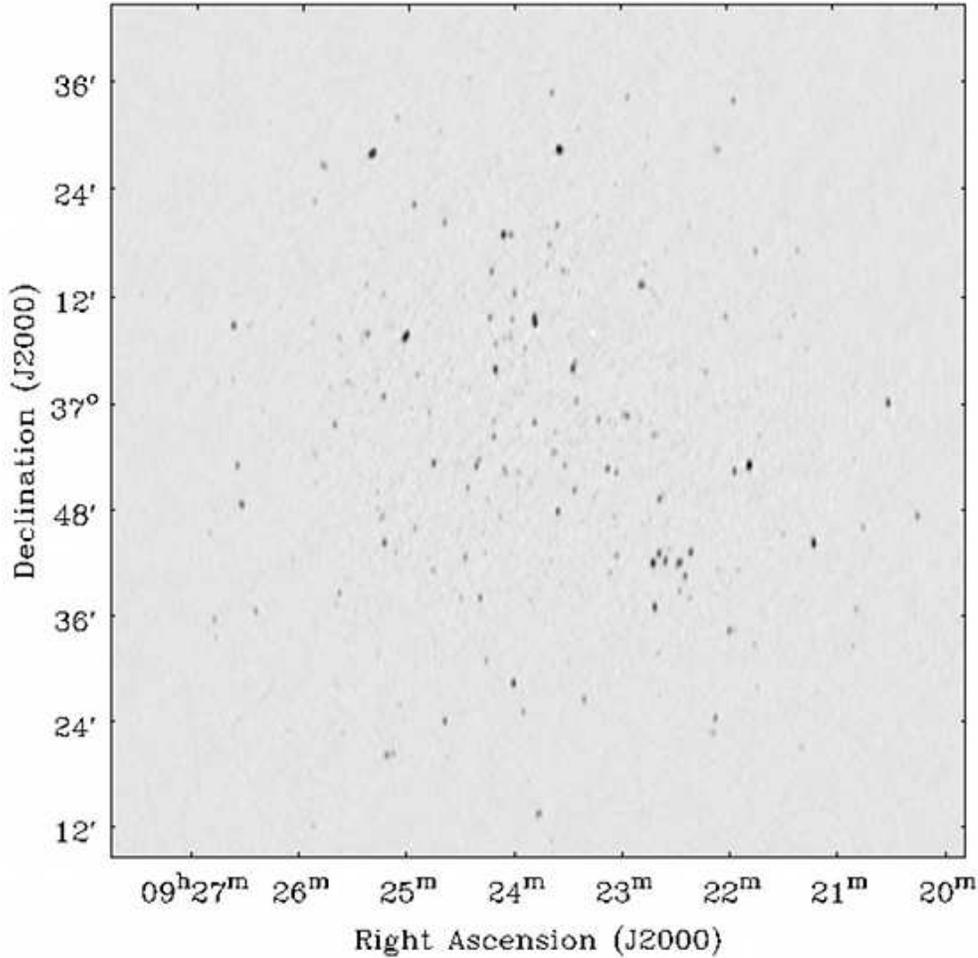}
 \caption{The radio continuum image of the CNOC2~0920+37 field at $z$~$\sim$~0.1 without applying primary beam correction. This image shows $\sim$~1.4 deg$^{2}$ of CNOC2~0920+37 and RMS noise level is about 12~$\mu$Jy~beam$^{-1}$.}
 \label{fig:cont}
\end{figure*}     

The observed data were reduced with both the Astronomical Image Processing System (AIPS) and the Common Astronomy Software Applications (CASA)\footnote{http://casa.nrao.edu} following standard procedures such as flagging, calibration, self-calibration and imaging. While AIPS was used to make the initial assessment of the observed data, discard corrupted data, and calibrate the observed data for flux and phase, CASA was used for self-calibration, imaging, and the subsequent analysis. The polarisation of the data products were first re-labeled to allow reduction of the linearly polarised data in AIPS (which assumes circular polarisation), after which the system temperature  (T$_{sys}$) correction was applied. We then manually edited the data contaminated by radio frequency interference by looking at the amplitudes in the frequency vs time domain for every baseline and every sub-band (IF). Calibration for flux and phase with 3C147 and 3C286 was performed along with the correction of the frequency-dependent gain variation of each antenna through a bandpass calibration procedure with the flux calibrator. After calibration, each sub-band of the data was processed separately in CASA. The data taken from 9 days were concatenated into separate sets for each of the 8 sub-bands. We self-calibrated each concatenated sub-band data with 6 bright radio continuum sources in the field to reduce residual amplitude and phase errors. In order to make continuum images of each sub-band for self-calibration, we used only the central 75~per~cent out of the 128 channels in each sub-band with pixel size of 5~arcsec per a pixel. The channels selected for continuum images were averaged into a new $uv$ data set consisting of 12 channels to avoid bandwidth smearing effects. Each channel of the new $uv$ data is an average of 8 channels of the original data. During self-calibration, three phase self-calibration loops and one amplitude self-calibration loop were repeated in each sub-band.
In the self-calibrated image of each sub-band, there remained artefacts caused by side lobes from bright off-axis sources. These were removed by a peeling technique as follows: Except for the target source to be peeled, all continuum sources were removed by subtraction of clean components from the $uv$ data, and then self-calibration was carried out with the target source. The gain tables taken from the self-calibration were applied to the original data and then the target source was subtracted from the visibility data. The final steps were to invert the source specific gain tables applied to the original data from which the target source was removed, and then add back the continuum clean components before moving on the next source whose artefacts required removal.

\begin{figure*}
 \includegraphics[trim=0mm 0mm 0mm 0mm, clip, width=150mm]{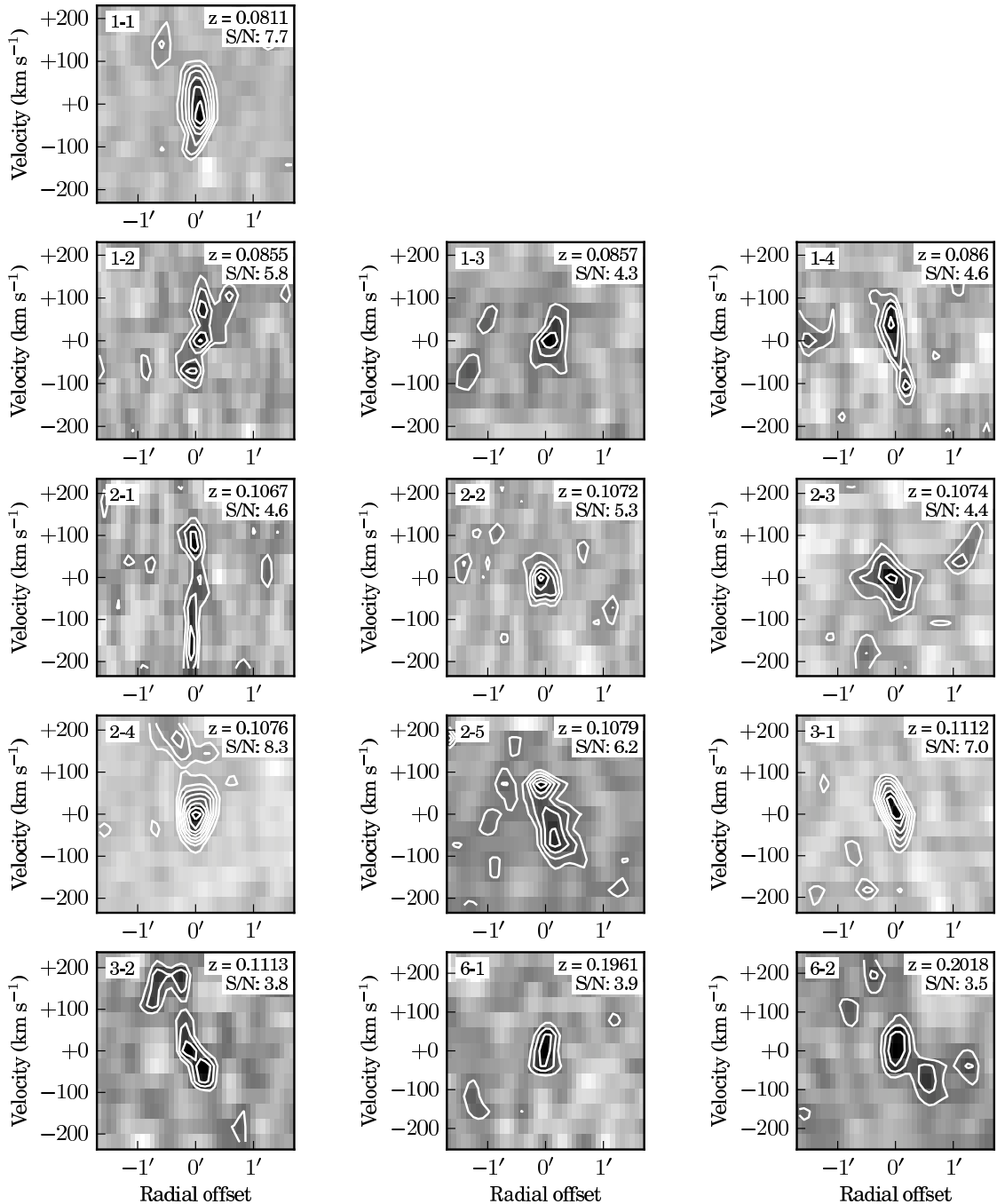}
 \caption{Position-velocity diagrams of 13 {\HI} detections made by {\sc duchamp}. The horizontal axis is radial offset from the central position of the {\HI} source along the major axis of the optical counterparts. It covers 3~arcmin on a side. The vertical velocity axis covers $\sim$400~km~s$^{-1}$ in the rest-frame of the galaxies. The contour levels are drawn at 1.5, 2.5, 3.5, 4.5, 5.5, 6.5, 7.5, 8.5 times the local noise ($\sigma_{\rm rms}$).}
 \label{fig:duchamp}
\end{figure*}

The final continuum images were made from the $uv$ data from which the confusing sources were peeled off. For construction of the continuum image for $z$~$\sim$~0.1, the $uv$ data from the first three sub-bands were concatenated. The 5 other sub-bands were used for the continuum image for $z$~$\sim$~0.2. When making the continuum images, we applied a multi-frequency deconvolution algorithm \citep{Rau_Cornwell:2011} in CASA (see Fig.~\ref{fig:cont}). The RMS of the continuum image at each redshift is $\sim$12~$\mu$Jy~beam$^{-1}$ and $\sim$13~$\mu$Jy~beam$^{-1}$, respectively. The astrometric accuracy  of the radio continuum sources was examined against their positions in the VLA FIRST survey. It is in good agreement within the maximum deviation of $\sim$3.5~arcsec (less than 1~pixel in the WSRT images). 

For the final spectral data cubes, the continuum sources were subtracted from the $uv$ data as follows. The continuum emission was deconvolved to obtain the clean component model of the continuum emission, and we subtracted the Fourier transform of the resulting model of the continuum directly from spectral $uv$ data using the CASA task `uvsub'. While most of continuum emission was removed, a small fraction of residual continuum flux remained. To remove the residual,  a first-order polynomial fit to the residual continuum was subtracted in the $uv$ domain using the CASA task `uvcontsub'. After subtracting the continuum flux, the final data cubes of each sub-band were made individually. The imaging step used a CASA weighting scheme of `briggs' with the robustness value of 0.5. The mean RMS noise level of spectral data cubes is $\sim$61~$\mu$Jy~beam$^{-1}$ per spectral channel and the typical synthesised beam size is $\sim$~33$\arcsec$~$\times$~20$\arcsec$.

\section{Measuring {\HI} 21-\lowercase{cm} emission}

\begin{table*}
 \caption{Properties of galaxies directly detected using {\sc duchamp}.}
 \label{tab:detection}
 \begin{tabular}{@{}ccccccccccc}
   \hline
      ID  & CNOC2 ID & SDSS ID & RA & Dec & S/N & $z_{\rm optical}$ & $z_{\rm radio}$ & $S_{\rm int}$ & $W_{\rm vel}$ &$M_{\HIsub}$ \\
           &  &  & (J2000) & (J2000) & & & & (mJy~km~s$^{-1}$) & (km~s$^{-1}$) &  (10$^9$~M$_{\sun}$) \\
      (1) & (2) & (3) & (4) & (5) & (6) & (7) & (8) & (9) & (10) & (11) \\ 
   \hline 
     1-1  & 80352 & 588297863107379261 & 09:24:36.10 & +37:03:09.29 & 7.7 & 0.0820 & 0.0811 & 32.6 & 179.4 & 0.97\\
     1-2  & 11165 & 588297863107313729 & 09:23:54.28 & +37:06:58.15 & 5.8 & 0.0867 & 0.0855 & 13.3 & 216.9 & 0.44\\
     1-3  & 92050 & 588297863107313874 & 09:24:20.04 & +37:01:25.40 & 4.3 & 0.0862 & 0.0857 & 4.3 & 72.3 & 0.14 \\
     1-4  & 131565 & 587734621636001861 & 09:22:59.76 & +37:08:23.26 & 4.6 & 0.0854 & 0.0860 & 13.4 & 180.9 & 0.45 \\
     2-1  & 101993 & 587735044691591273 & 09:24:20.89 & +36:53:39.54 & 4.6 & 0.1067 & 0.1067 & 10.7 & 295.2 & 0.55\\
     2-2  & 161711 & 588297863107117112 & 09:22:20.45 & +36:49:48.45 & 5.3 & 0.1073 & 0.1072 & 26.7 & 73.9 & 1.40 \\
     2-3  & 121374 & 588297863107313705 & 09:23:57.28 & +36:58:17.32 & 4.4 & 0.1077 & 0.1074 & 6.3 & 73.9 & 0.33 \\
     2-4  & 141529 & 588297863107182723 & 09:22:49.32 & +36:59:31.53 & 8.3 & 0.1085 & 0.1076 & 39.2 & 333.0 & 2.07 \\
     2-5  & 91661   & 588297863107313700 & 09:24:28.38 & +36:59:52.20 & 6.2 & 0.1079 & 0.1079 & 20.9 & 148.0 & 1.11 \\
     3-1  & -          & 587735241716138078 & 09:24:53.36 & +36:46:30.54 & 7.0 & 0.1110 & 0.1112 & 49. 7 & 110.3 & 2.81 \\
     3-2  & -          & 587735044691460246 & 09:22:55.60 & +36:41:14.09 & 3.8 & 0.1114 & 0.1113 & 18.9 & 257.6 & 1.07 \\
     6-1   & 140493 & 588297863107248248 & 09:23:11.54 & +36:55:11.71 & 3.9 & 0.1967 & 0.1961 & 6.4 & 79.6 & 1.17 \\
     6-2   & 120605 & 588297863107248267 & 09:23:35.67 & +36:55:18.42 &  3.5 & 0.2022 & 0.2018 & 2.4 & 80.3 & 0.47 \\
   \hline
  \end{tabular}
\end{table*}

\subsection{Direct detection}
The eight data cubes were searched for direct detections of {\HI} line emission from individual galaxies using {\sc duchamp}\footnote{http://www.atnf.csiro.au/people/Matthew.Whiting/Duchamp/}, a source finding software package \citep{Whiting:2012}. Cross-checking the {\sc duchamp} detections in the low-redshift  ($z$~$\sim$~0.1) data cubes with the CNOC2~0920+37 catalogue as well as the optical counterpart from the SDSS revealed that 9 objects detected by {\sc duchamp} matched those in the CNOC2~0920+37 sample. Additionally, two objects were confirmed as detections, the SDSS providing spectroscopic redshifts for them. A total of 11 objects were detected with optical counterparts of galaxies at $z$~$\sim$~0.1 by {\sc duchamp}. The two newly found galaxies were not listed in the CNOC2~0920+37 catalogue because they are located out of the range of the CNOC2 survey area although they lie within the WSRT primary beam coverage. The objects confirmed as true detections have detection threshold of more than 4$\sigma$. In the higher redshift cubes at $z$~$\sim$~0.2, only two objects were detected with 3~$<$~$\sigma$~$<$~4, and the very highest redshift where we made an individual detection of a spectroscopically identified object was $z=$~0.2022. 

Position-velocity diagrams of all 13 individual detections are plotted in Fig.~\ref{fig:duchamp}. The position angle is taken to be the same as that of the optical counterpart to the {\sc duchamp} detection. We note that some (but not all) of the position-velocity diagrams show velocity gradients consistent with the double peaked velocity profiles typical of late-type galaxies, which means that some galaxies are slightly resolved even with the large synthesised beam of the WSRT ($\sim$~33$\arcsec$~$\times$~20$\arcsec$). Velocity gradients are not detected in others, presumably because they are not spatially resolved by the WSRT synthesised beam. The influence of the synthesised beam width of the WSRT on the stacked signal is discussed in the next section. 

Table~\ref{tab:detection} lists all basic properties of direct detections, measured and derived by {\sc duchamp}. The detailed description of each column is the following: (1) assigned ID number based on combination of the number of data cube and the assigned number in increasing redshift where each detection was made; (2) CNOC2~0920+37 catalogue ID matched with direct detections; (3) SDSS ID of counterpart of direct detections; (4) (5)  coordinates of direct detections; (6) signal-to-noise ratio at the peak flux; (7) redshift determined by optical spectroscopic observation of CNOC2 and SDSS; (8) redshift derived from {\sc duchamp} detection; (9) integrated {\HI} flux with the primary beam correction applied; (10) the full velocity width of the detections; (11) {\HI} mass estimated from integrated flux (column 9) with correcting the primary beam attenuation effect.
   
Note that these individual detections are not treated separately in the subsequent stacking analysis (Section~3.2), that is, the {\HI} mass and flux measurements in Table~\ref{tab:detection} are not used for co-adding {\HI} flux and calculating {\HI} mass. However, they are included in the list to be stacked and do contribute to the integral {\HI} signal when the spectra for the CNOC2~0920+37 sample are stacked. The two newly detected non-CNOC2 objects are excluded from the following analysis.

\subsection{{\HI} stacking}
The previous section has shown that only about 7 per cent (11/155) of the optically selected galaxies in our sample are individually detected at higher than 4$\sigma$ and 3$\sigma$ significance at $z$~$\sim$~0.1 and 0.2, respectively. Therefore we do not attempt to calculate {\OHI} based on these individual detections, but instead we make use of a spectral signal stacking technique to measure the {\HI}  content of galaxies at $z$~$\sim$~0.1 and 0.2. The final three-dimensional data cube of each sub-band has two position axes (R.A. and Dec.) of dimension 1$\degr$~$\times$~1$\degr$ and one frequency axis with 128 frequency channels, equivalent to $\sim$4800~km~s$^{-1}$ in velocity depth. Since redshifts from the CNOC2~0920+37 catalogue can be converted to the corresponding frequencies in data cubes, we can determine the expected positions of the {\HI} 21-cm emission in the data cube and extract spectra at those locations. Before stacking {\HI} emission lines, the primary beam correction should be considered. The primary beam response pattern of the WSRT varies with distance from the pointing centre and observing frequency. We apply the primary beam correction using the primary beam attenuation function, \begin{math} {g(\nu,r)={\rm cos^6}(c \times \nu \times r)} \end{math}, where $r$ is the distance from the pointing centre in degrees, $\nu$ is observing frequency in GHz and $c$ is a constant ($c=$~68 in our observing frequency band)\footnote{http://www.astron.nl/radio-observatory/astronomers/wsrt-guide-observations/5-technical-information/5-technical-informatio}. This means that for each target the extracted spectrum ($S_{i}$) and RMS noise ($\sigma_{i}$) are divided by the the primary beam attenuation function, $g_{i}(\nu,r)$:
\begin{equation}
  S_{i}' = S_i/g_i, \qquad \sigma_{i}' = \sigma_i/g_i.
\label{eq:pb}
\end{equation}
 After this correction spectra are shifted to the rest-frame velocity of each galaxy and the {\HI} emission spectra are co-added to achieve a low-noise average 21-cm {\HI} spectrum of the whole sample, based on the classified galaxy types.

\begin{figure}
 \centering
 \includegraphics[trim=0mm 0mm 0mm 0mm, clip, width=84mm]{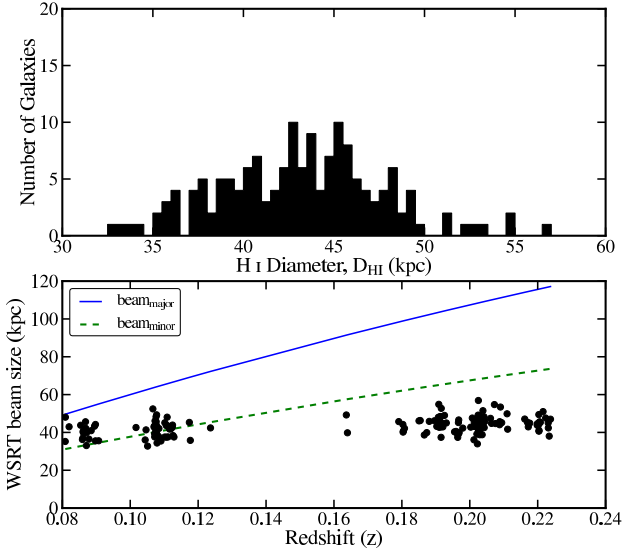}
 \caption{The distribution of estimated {\HI} diameter of galaxies in the CNOC2~0920+37 catalogue (in upper panel) and the WSRT beam size in unit of kpc as a function of redshift (in lower panel). The derived {\HI} diameters of the CNOC2~0920+37 galaxies (filled black circles) are compared with WSRT beam size in the lower panel. Many galaxies in the $z \sim 0.1$ subsample are likely to be slightly resolved in the east-west direction.}
 \label{fig:galaxysize}
\end{figure}

The synthesised beam size of the WSRT is $\sim$33$\arcsec$~$\times$~20$\arcsec$, corresponding to $\sim$50~kpc~$\times$~31~kpc in the data cube with lowest redshifts and increasing in size toward higher redshifts. 
In order to check whether galaxies in the data cube are resolved by the WSRT synthesised beam, we estimate the {\HI} size of galaxies in our sample using the relation between optical radius and {\HI} diameter derived from local spiral and irregular galaxies by \citet{Broeils_Rhee:1997}. They found that there is a strong correlation between the diameter of the optical isophote (D$_{25}$) in $B$-band and the diameter (D$_{\HIsub}$) of the contour where the atomic hydrogen surface density has dropped to 1 M$_{\sun}$~pc$^{-2}$. Since we do not have information about the optical $B$-band diameter of our sample galaxies, we instead used R$_{90}$, the Petrosian radius, containing 90 per cent of the Petrosian $r$-band flux, as listed in the SDSS. We also use the SDSS data to determine the Petrosian radii of the galaxies in the \citet{Broeils_Rhee:1997} sample, and thus obtained a relation between the {\HI} and the Petrosian radius. From the newly derived relation, we estimate the {\HI} size of the sample galaxies. As seen in Fig.~\ref{fig:galaxysize}, most galaxies have D$_{\HIsub}$ less than 50~kpc, but a few are in the range of 50 to 57~kpc. The implication in Fig.~\ref{fig:galaxysize} that some of the galaxies in the low redshift sample spill outside of the WSRT synthesised beam is consistent with the detection of rotational velocity gradients revealed in Fig.~\ref{fig:duchamp}. In the higher redshift data cubes, the galaxy {\HI} signal is expected to be captured by the full resolution beam size. 

In order to test the success of the WSRT beam at including the total {\HI} signal for all galaxies, additional sets of image cubes were constructed with larger beams with half-power widths of $\sim$~33$\arcsec$~$\times$~33$\arcsec$, 45$\arcsec$~$\times$~45$\arcsec$, and 60$\arcsec$~$\times$~60$\arcsec$. Along with the default resolution cubes, these lower resolution image cubes were carried through the stacking procedure as described below.

\begin{figure}
 \centering
 \includegraphics[trim=0mm 0mm 0mm 0mm, clip, width=82mm]{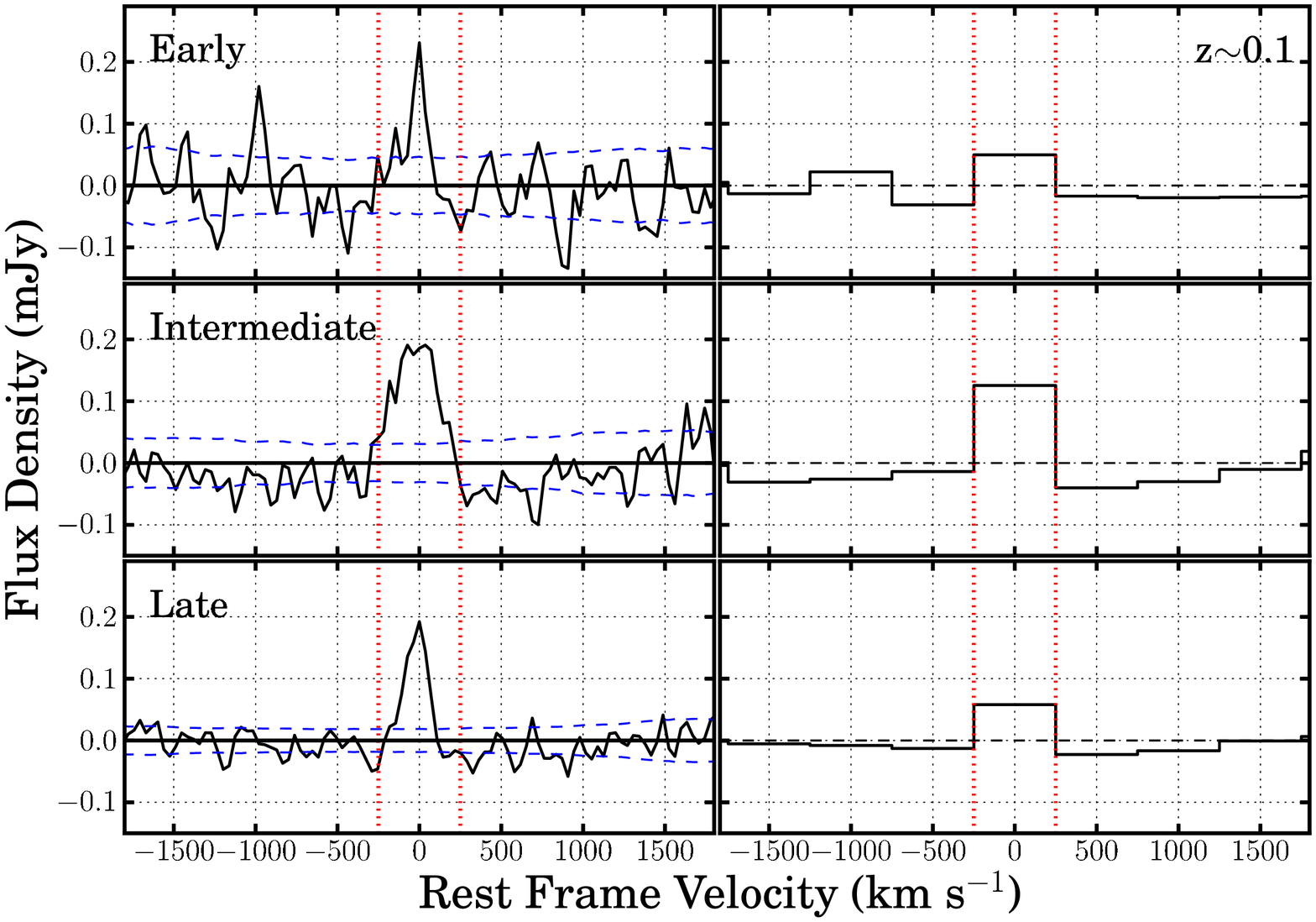}
 \includegraphics[trim=0mm 0mm 0mm 0mm, clip, width=82mm]{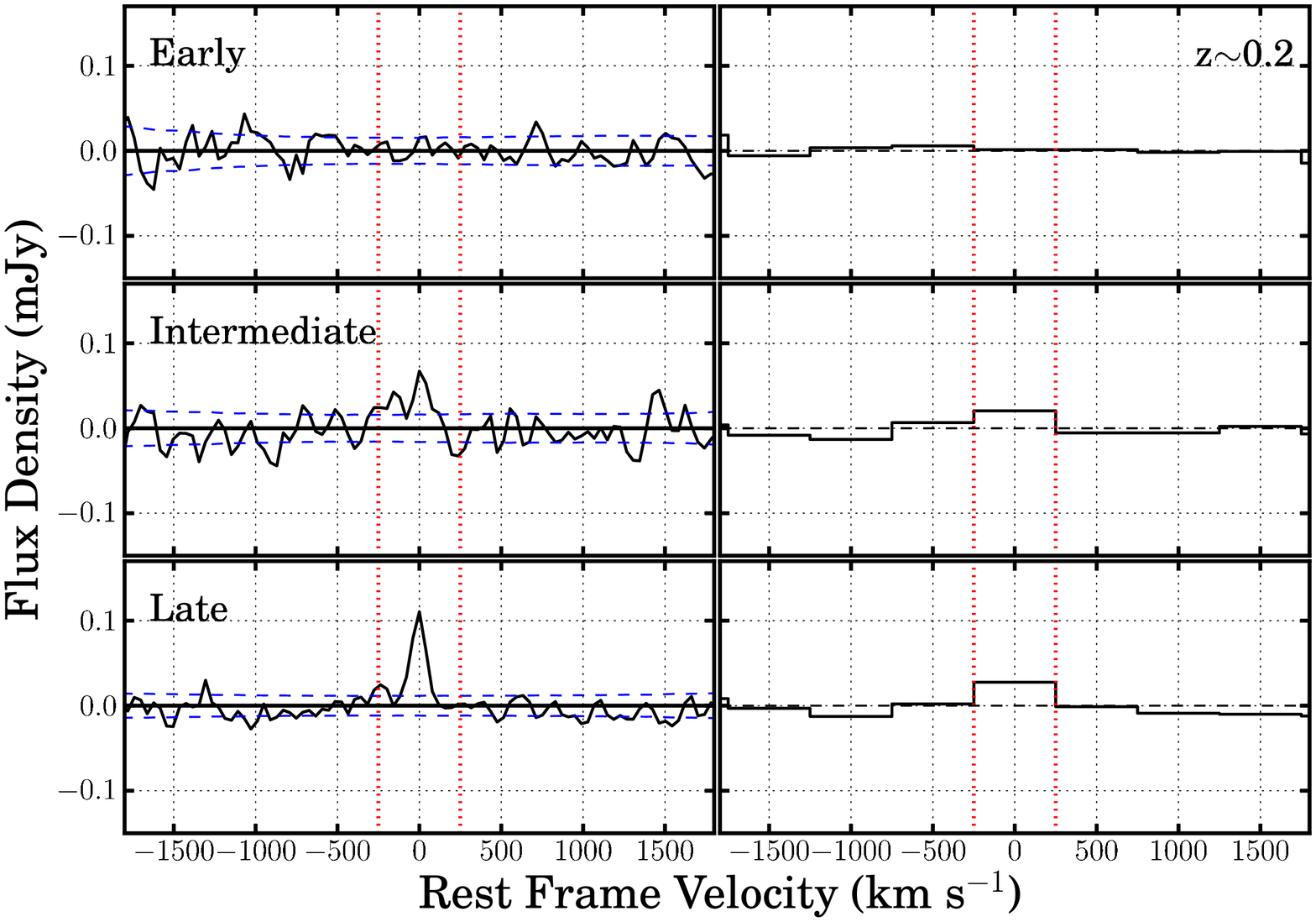}
 \caption{The averaged {\HI} emission spectra of galaxies in each subgroup after stacking. The top panels are the co-added {\HI} spectra at $z$~$\sim$~0.1, for the different galaxy types, and the  bottom panels are for galaxies at $z$~$\sim$~0.2. The vertical dashed line in each panel shows the velocity width within which all {\HI} emission from galaxies is included. The 1$\sigma$ error is shown as a horizontal dashed line in each panel. To the right of each panel are the same spectra but re-binned to a velocity width of 500 km~s$^{-1}$.}
 \label{fig:stack}
\end{figure}

\begin{table*}
 \caption{{\HI} mass measurements ({\MHI}), $B$-band luminosity density ($\rho_{L_{B}}$) and {\HI} mass density ({\rhoHI}) of each galaxy type at the redshift of $z$~$\sim$~0.1 and 0.2, respectively.}
 \label{allmeasurement}
 \begin{tabular}{@{}lcccccccccccc}
   \hline
                & \multicolumn{5}{c}{$z$~$\sim$~0.1} & & \multicolumn{5}{c}{$z$~$\sim$~0.2}  \\
   \cline{2-6} \cline{8-12} 
    Sample  & $N_{\rm gal}$ & $\langle \MHI \rangle$ & $\langle L_B \rangle$ & $\rho_{L_{B}}$ & $\rhoHI$ & & $N_{\rm gal}$ & $\langle \MHI \rangle$ & $\langle L_B \rangle$ & $\rho_{L_{B}}$ & $\rhoHI$ \\
   \hline 
     Early & 8 & 1.18~$\pm$~0.39 & 7.85 & 6.45~$\pm$~1.12 & 0.79~$\pm$~0.30 & & 25 & 0.13~$\pm$~0.46 & 16.35 & 6.48~$\pm$~1.35 & 0.05~$\pm$~0.18 \\
     Intermediate & 17 & 3.11~$\pm$~0.45 & 14.13 & 4.42~$\pm$~0.88 & 1.03~$\pm$~0.25 & & 25 & 1.94~$\pm$~0.52 & 13.29 & 4.93~$\pm$~1.26 & 0.71~$\pm$~0.26\\
     Late & 34 & 1.43~$\pm$~0.20 & 4.41 & 6.09~$\pm$~1.19 & 2.61~$\pm$~0.63 & & 46 & 2.61~$\pm$~0.35 & 9.15 & 7.90~$\pm$~2.18 & 3.81~$\pm$~1.16 \\
   \hline
     All                & & \multicolumn{4}{c}{{\OHI}($z$~$\sim$~0.1)~$=$~(0.33~$\pm$~0.05)~$\times$~10$^{-3}$} & & &  \multicolumn{4}{c}{{\OHI}($z$~$\sim$~0.2)~$=$~(0.34~$\pm$~0.09))~$\times$~10$^{-3}$} \\
   \hline
 \end{tabular}

 \medskip 
Note: $N_{\rm gal}$ is the number of galaxies that are co-added and $\langle \MHI \rangle$ is averaged {\HI} mass (in units of 10$^{9}$~M$_{\sun}$), $\langle L_B \rangle$ is mean $B$-band luminosity in units of 10$^{9}$~L$_{\sun}$, $\rho_{L_{B}}$ the luminosity density in units of 10$^{7}$~L$_{\sun}$~Mpc$^{-3}$ and $\rhoHI$ the {\HI} density in units of 10$^{7}$~M$_{\sun}$~Mpc$^{-3}$.
 \label{tab:allmeasurement}
\end{table*}

When stacking galaxy spectra, a weighted average is made. The spectra to be co-added using a weighted average scheme have been corrected with the primary beam attenuation function as in equation (\ref{eq:pb}). The per channel weight is computed from the image plane RMS, corrected by the primary beam correction factor at the position of the galaxy. The formal average specification is given by:
\begin{equation}
  S_{\rm average} = \frac{\sum_i w_{i}' S_{i}'}{\sum_i w_{i}'}, \qquad w_{i}' = \frac{1}{{\sigma_{i}'}^2},
\end{equation}
where $S_{i}'$ is the primary beam corrected spectrum. The RMS noise, $\sigma_{i}'$, is measured per frequency channel in each data cube with applying the primary beam correction and the inverse of the square of the RMS noise is applied during the stacking procedure as a statistical weight, $w_{i}'$. 

Recall (see Fig.~\ref{fig:hist}) that the sample was divided into two sub-samples at $z\sim$~0.1 and $z\sim$~0.2. The first sub-sample consists of 59 galaxies lying in the first three sub-bands, i.e. in the frequency range 1321~MHz to 1263~MHz. The remaining 96 galaxies lie in the last five sub-bands i.e. in the frequency range 1222~MHz to 1160~MHz. These sub-samples at different redshifts are each further divided into three morphological types (see Fig.~\ref{fig:cmd}). Fig.~\ref{fig:stack} shows the results of the {\HI} spectra stacking for each sub-sample separately. Five of the six stacked spectra show a clear emission line signal, in particularly for the intermediate and late-type galaxies.

From the stacked spectrum, one can determine the co-added flux. This co-added {\HI} flux can be converted to {\HI} mass using the equation \citep{Wieringa:1992}:
\begin{equation}
  \frac{{\MHI}}{\rmn{M_{\sun}}} = \frac{236}{(1+z)} \left( \frac{D_L}{\rmn{Mpc}} \right)^2 
  \bigg( \frac{\int S_V dV}{\rmn{mJy \,km~s^{-1}}} \bigg)
\end{equation} 
where $\int S_V dV$ is the integrated {\HI} emission flux in units of $\rmn{mJy \,km~s^{-1}}$ and $D_{L}$ is the luminosity distance in unit of Mpc. For the calculation of $\int S_V dV$ it is assumed that a velocity range $\Delta V$ (specified in the rest frame of the {\HI} emitter) of 500~km~s$^{-1}$  contains all the {\HI} emission. This assumption takes into account both the average {\HI} velocity width of the sample galaxies and the uncertainty in the optical redshifts. For the HIPASS Bright Galaxy Catalogue \citep[BGC,][]{Koribalski:2004}, a catalogue of the 1000 {\HI} bright galaxies in the local Universe, the mean velocity width $w_{20}$ is 210~km~s$^{-1}$ ($\sigma$~$=$~106~km~s$^{-1}$). This is consistent with the mean velocity width of the CNOC2~0920+37 sample galaxies, derived from the Tully-Fisher relation \citep{Tully:1977}. Using this relation between absolute $B$-band magnitude and velocity width ($w_{20}$) \citep{Tully:2000} with no inclination correction, the averaged velocity width of $\sim$~240~km~s$^{-1}$ ($\sigma$~$\sim$~82~km~s$^{-1}$) is estimated. The expected velocity width of the highest luminosity galaxies in the sample is $w_{20}=$~560~km~s$^{-1}$. The statistical uncertainty of the CNOC2 optical redshifts is approximately 79~km~s$^{-1}$. Taking into account the redshift uncertainty, a velocity width of 500~km~s$^{-1}$ was used to measure all {\HI} signals from all co-added galaxies. 

We estimate the error of the {\HI} mass measurement through jacknife resampling \citep{Efron:1982}. From a sample of {\HI} spectra ($X = (x_1,x_2,...,x_n $)) to be co-added, one spectrum is removed at a time to construct as many jacknife samples as the number of samples: 

\begin{equation}
  X_i = (x_1,x_2,...,x_{i-1},x_{i+1},x_n)
\end{equation} 
for $i$~=~1, 2, $...$, $n$. With $i$th jacknife sample, $i$th partial co-added spectrum is obtained 
\begin{equation}
  \hat{\alpha_i} = f(X_i).
\end{equation} 
Then, the pseudo-values are defined as 
\begin{equation}
  \hat{\alpha_{i}^{*}} = n\hat{\alpha} - (n-1)\hat{\alpha_i}
\end{equation} 
where $\hat{\alpha}$ is the spectrum co-added without jacknife resampling. The error of the stacked {\HI} spectrum can be expressed as  
\begin{equation}
\label{eq:error}
   (\sigma^{*})^2 = \frac{1}{n(n-1)} \sum_i{(\hat{\alpha_{i}^{*}}-\hat{\alpha^{*}})^2}, \quad \hat{\alpha^{*}} = \frac{1}{n} \sum_i^n{\hat{\alpha_{i}^{*}}}.
\end{equation}

This error is propagated in the conversions to {\HI} mass and density. Table~\ref{allmeasurement} shows measurements of {\HI} mass and error derived from the above relations as a function of galaxy type at $z\sim$~0.1 and 0.2, respectively.

Comparison of the results of stacked {\HI} signal strength as a function of the size of the synthesised beam revealed a monotonic increase in the signal strength (summed over all galaxy types) as the beam area was increased for both $z$~$\sim$~0.1 and 0.2. For the low redshift subsample, this amounted to an increase of 7 per~cent in integrated {\HI} mass as the beam was increased from 56~$\times$~35~kpc to 56~$\times$~56~kpc. For the high redshift subsample, increasing the area from 112~$\times$~71~kpc to 112~$\times$~112~kpc  produced a 2 per~cent increase. Further doubling of the beam diameter led to an additional $\sim$11 per~cent for both samples, but measurements on these distance scales are vulnerable to contamination by companion and group galaxies. To adequately capture the total {\HI} mass contribution from each target galaxy while avoiding contamination, we choose to use the 56~$\times$~56~kpc beam at low redshift and the 112~$\times$~71~kpc beam for high redshift. The measurements in Table~\ref{allmeasurement} are based on these beam sizes. 
This adjustment to the low redshift sample is less than the statistical uncertainties calculated in equation~(\ref{eq:error}), but it avoids a possible bias due to galaxies spilling out of the higher resolution beam.

\begin{figure}
 \centering
 \includegraphics[trim=0mm 0mm 0mm 0mm, clip, width=84mm]{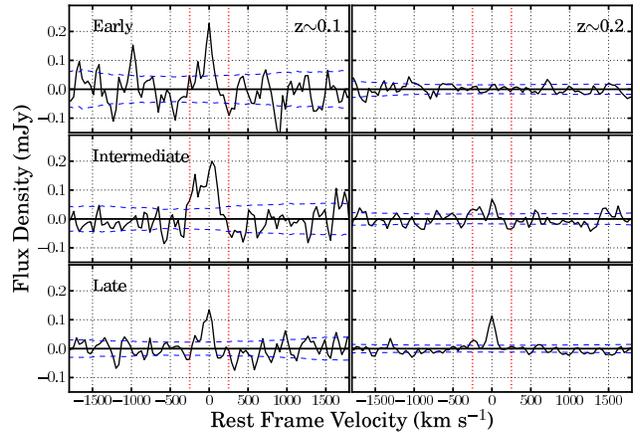}
 \caption{The stacked {\HI} emission spectra of galaxies in each subgroup, exclusive of objects directly detected by {\sc duchamp}. The left panels are the co-added {\HI} spectra for the different galaxy types at $z$~$\sim$~0.1, and the right panels are for galaxies at $z$~$\sim$~0.2. The vertical dashed line in each panel shows the velocity width (500 km~s$^{-1}$). The 1$\sigma$ error is shown as a horizontal dashed line in each panel.}
 \label{fig:stack_no_detection}
\end{figure}

\begin{table}
 \caption{{\HI} mass measurements in units of 10$^{9}$~M$_{\sun}$ excluding direct detections, $z$~$\sim$~0.1 and 0.2, respectively.}
 \centering
 \label{tab:HImass_no_detection}
 \begin{tabular}{@{}lccc}
   \hline
                    & $z$~$\sim$~0.1 & $z$~$\sim$~0.2  \\
       Sample  & $\langle \MHI \rangle$ & $\langle \MHI \rangle$ \\
   \hline 
     Early & 1.18~$\pm$~0.39 & 0.13~$\pm$~0.46 \\
     Intermediate & 2.36~$\pm$~0.48 & 1.37~$\pm$~0.50 \\
     Late & 0.80~$\pm$~0.17 & 2.63~$\pm$~0.34 \\
   \hline
 \end{tabular}
\end{table}

To see to what extent directly detected objects contribute to the stacked {\HI} mass, we reproduce the co-added {\HI} emission spectra, excluding the 11 direct detections from the CNOC2~0920+37 catalogue (see Fig.~\ref{fig:stack_no_detection}) and compare the {\HI} mass measurements with those calculated from the averaged {\HI} spectra including all galaxies, as in Table~\ref{allmeasurement}. At both redshift bins, all direct detections are intermediate-type or late-type galaxies. 6 late-type galaxies and 3 intermediate-type galaxies were detected at $z \sim$~0.1, and 2 galaxies (1 intermediate-type and 1 late-type) were detected at $z \sim$~0.2, using {\sc duchamp}. As seen in Table~\ref{tab:HImass_no_detection}, after removing direct detections from the stacking list, the {\HI} mass decreases
24~per~cent and 44~per~cent in intermediate-type galaxies and late-type galaxies at $z \sim$~0.1, respectively. At $z \sim$~0.2, the {\HI} mass of intermediate galaxies decreases by 29~per~cent, while there is no change in the  {\HI} mass of late-type galaxies. In spite of the relatively small number of direct detections, individually detected objects significantly contribute to the co-added {\HI} mass.

\section{The Cosmic {\HI} Gas Density ({\OHI})}

To see how the {\HI} gas content of galaxies has varied with the cosmic time, we calculate {\OHI}, the cosmological {\HI} mass density, expressed as a fraction of the present critical density. Calculating {\OHI} requires an accurate volume normalisation, which in our approach is dependent on good measurements of the luminosities and the total luminosity density of the CNOC2 galaxies that are used for the {\HI} stacking. 

We make use of the optical luminosity density of the CNOC2~0920+37 field as a function of galaxy type from \citet{Lin:1999}. They investigated the evolution of the luminosity function with redshift and galaxy type and provide luminosity densities $\rho_{L_{B}}(\it z)$ up to $z$~$\sim$~0.55. Their galaxy classification scheme is the same as the one we used. The $B$-band luminosity and luminosity density of each galaxy type are used in the conversion of {\HI} gas density as follows:  
\begin{equation}
  {\rhoHI} = \frac{\sum \MHI}{\sum L_B} \times \rho_{L_{B}(\it z)}, \qquad 
\sum \MHI = \langle \MHI \rangle \times N
 \label{eq:rhohi}
\end{equation} 
where $N$ is the number of co-added galaxies, $L_{B}$ is luminosity in the $B$-band taken from the CNOC2~0920+37 catalogue, and $\langle \MHI \rangle$ denotes the averaged {\HI} mass we measured using the stacking technique. We separately calculate {\HI} gas density with respect to each galaxy type of early, intermediate and late-type at each redshift bin (see Table~\ref{allmeasurement}). 
It should be noted that this method of evaluating {\rhoHI} intrinsically assumes that there is no strong variation of relative gas richness over the total range of $L_B$ of the total galaxy population in each subsample. In practice, we can only measure the gas content of galaxies above the completeness limit imposed by the CNOC2 survey. If the relative gas richness of galaxies below the optical detection limit is different, we will derive a biased measurement of {\rhoHI}. In fact, local studies show that there is a correlation between gas richness and optical luminosity. We make use of these local correlations to estimate a correction factor $f$ that we apply to equation (\ref{eq:rhohi}), such that our measurement of {\rhoHI} represents the gas content of the total galaxy population, rather than that of the galaxies above our detection limit. We assume that the relations between gas richness and luminosity have not changed significantly between $z=$~0 and $z=$~0.2. We refer to Appendix A for a detailed derivation of $f$. The total {\HI} gas density in the low and high redshift bins in Table~\ref{allmeasurement} are obtained by adding the {\HI} density values for the three galaxy types. 

Note that in order to derive luminosity densities, \citet{Lin:1999} applied different cosmological parameters ($\Omega_{\Lambda}=$~0.0, $\Omega_{M}=$~1.0 and $H_{0}=$~100~km~s$^{-1}$~Mpc$^{-1}$) from the current concordance $\Lambda$CDM cosmology as adopted in this paper ($\Omega_{\Lambda}=$~0.7, $\Omega_{M}=$~0.3 and $H_{0}=$~70~km~s$^{-1}$~Mpc$^{-1}$). The correction for the different cosmological parameters is made using the correction factor inferred from the ratio of $D_L^2 / D_C^3$ between two cosmological parameter sets because {\HI} density is proportional to $D_L^2 / D_C^3$, where $D_L$ and $D_C$ are luminosity distance and co-moving distance, respectively. The correction factor amounts to 0.667 and 0.640 for the low and high redshift samples, respectively. The corrected luminosity densities are provided in Table~\ref{allmeasurement}. Given the corrections, we obtain the {\HI} mass density of early, intermediate and late-type galaxies in each redshift bin, and these are also listed in Table~\ref{allmeasurement}. The errors on the {\HI} mass density take into account both the statistical uncertainty on the stacked {\HI} mass measurement and the measurement error on the $B$-band luminosity density. 

\begin{figure*}
 \includegraphics[trim=0mm 0mm 0mm 0mm, clip, width=180mm]{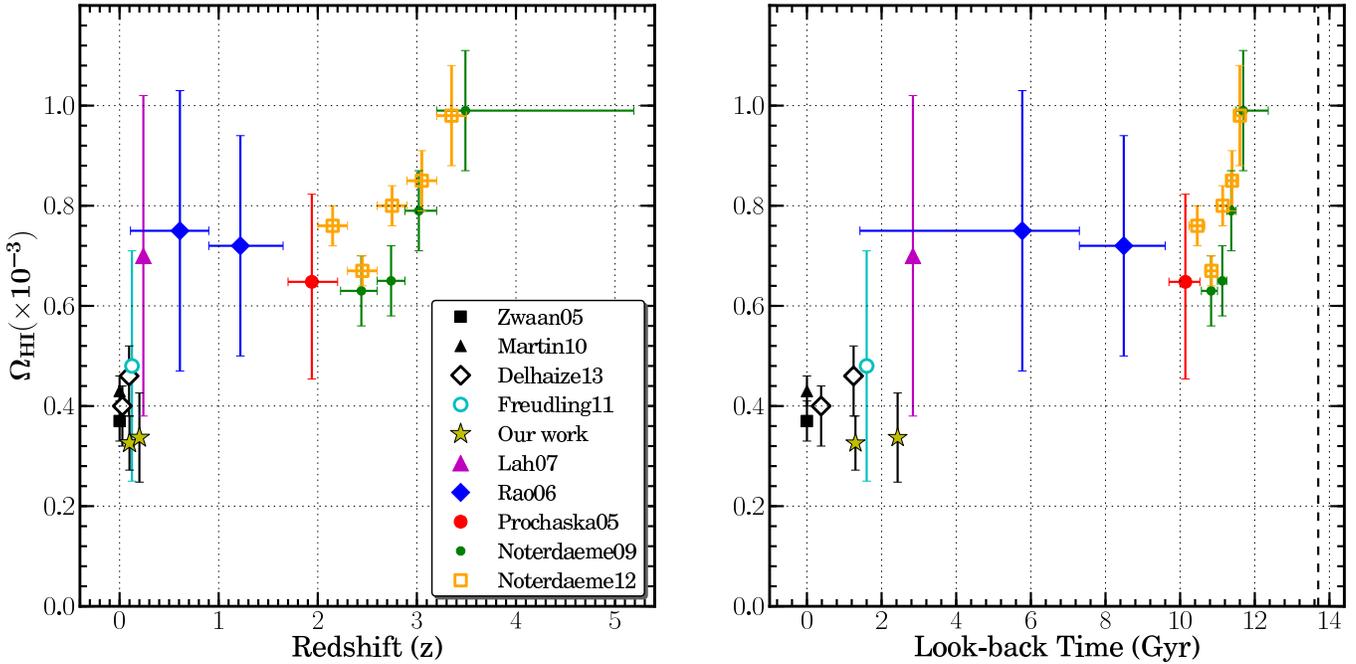}
 \caption{The cosmic {\HI} gas density ({\OHI}) evolution plots, inclusive of our measurements, as a function of redshift and look-back time, respectively. All measurements are made with the same cosmological parameters. The small black square is the HIPASS 21-cm emission measurement by \citet{Zwaan:2005}. The small black triangle is the ALFALFA 21-cm emission measurement from \citet{Martin:2010}. The open diamonds are measurements from the HIPASS and the Parkes observation of the SGP field using a {\HI} stacking technique \citep{Delhaize:2013}. The cyan open circle is the preliminary result from the AUDS \citep{Freudling:2011}. The big purple triangle is the measurement from the GMRT 21-cm emission stacking \citep{Lah:2007}. The blue diamonds, red big circle, green circles, and orange open squares are damped Lyman-$\alpha$ measurements from the HST and the SDSS by \citet{Rao:2006}, \citet{Prochaska:2005}, \citet{Noterdaeme:2009}, \citet{Noterdaeme:2012}, respectively. Our measurements at $z$~$\sim$~0.1 and 0.2 are the yellow stars. In the right panel, the vertical dashed line represents the age of the Universe.}
 \label{fig:omegaHI}
\end{figure*}  

The cosmic {\HI} gas density is then calculated as {\HI} gas density divided by the critical density ($\rho_{\rm crit}$): 
\begin{equation}
  {\OHI} = \frac{\rhoHI}{\rho_{\rm crit}}, \quad \rho_{\rm crit} = \frac{3H_{0}^{2}}{8\pi G},
 \label{eq:ohi}
\end{equation} 
where $H_{0}$ is the Hubble constant and $G$ is the gravitational constant. From equation~(\ref{eq:ohi}) we obtain 
{\OHI}~$=$~(0.33~$\pm$~0.05)~$\times$~10$^{-3}$ and 
{\OHI}~$=$~(0.34~$\pm$~0.09)~$\times$~10$^{-3}$ at $z$~$\sim$~0.1 and $z$~$\sim$~0.2, respectively, for all galaxy types summed.

\begin{table}
 \caption{The {\HI} mass to luminosity ratio ($\MHI/L_B$) as a function of galaxy types of CNOC2~0920+37 sample at the redshift of $z$~$\sim$~0.1 and 0.2, respectively in the unit of $\rmn{M_{\sun}/L_{\sun}}$.}
 \label{ML}
 \centering
 \begin{tabular}{@{}lccc}
   \hline
                & $z$~$\sim$~0.1 & & $z$~$\sim$~0.2  \\
    Sample  & $\langle \MHI \rangle / \langle L_B \rangle$ & & $\langle\MHI \rangle / \langle L_B \rangle$ \\
   \hline 
     Early & 0.15~$\pm$~0.05& & 0.01~$\pm$~0.03 \\
     Intermediate & 0.22~$\pm$~0.03 &  & 0.15~$\pm$~0.04 \\
     Late & 0.32~$\pm$~0.04 & &0.29~$\pm$~0.04 \\
   \hline 
 \end{tabular}
\end{table}

At $z$~$\sim$~0.2, it is found that most of the {\HI} gas resides in intermediate and late-type galaxies and the fraction of {\HI} gas in early-type galaxies is less than 2 per~cent. In contrast,  at $z$~$\sim$~0.1 early-type galaxies contribute approximately 22 per~cent to {\OHI}. The reason for the higher {\HI} mass density of early-type galaxies at $z$~$\sim$~0.1 might be a sensitivity effect. Gas richness  ${\MHI}/{L_B}$ is a function of luminosity in the sense that higher luminosity galaxies have lower values of ${\MHI}/{L_B}$ \citep{Roberts:1994} and this relation is stronger for early-type galaxies. As seen in Table~\ref{ML}, early-type galaxies at $z$~$\sim$~0.1 have higher mean ${\MHI}/{L_B}$ ratio than those in $z$~$\sim$~0.2 and the ratio is also higher than that \citet{Roberts:1994} found for early-type galaxies (${\MHI}/{L_B} \la 0.1$).  Thus, while gas-rich and low luminosity early-type galaxies might be included in the lower redshift bin, the higher redshift bin might only include gas-poor and higher luminosity early-types. The other possibility to explain the high ${\MHI}/{L_B}$ of early-type galaxies at $z \sim$~0.1 is the contamination of galaxy classification from other galaxy types. As stated in Section 2.1, we adopted the same galaxy classification scheme used by \citet{Lin:1999}. They warned that their classified galaxy types may have biases. This means that some of those classified as early-type galaxies in the CNOC2 catalogue could be other morphological types such as lenticular or spiral galaxies. Furthermore, due to the small number of early-type galaxies to be stacked, one or two galaxies incorrectly classified can significantly affect ${\MHI}/{L_B}$. 

Fig.~\ref{fig:omegaHI} shows {\HI} gas density values as a function of redshift and look-back time from our study as well as the previous studies by \citet{Zwaan:2005,Martin:2010,Freudling:2011,Lah:2007,Rao:2006,Prochaska:2005,Noterdaeme:2009,Noterdaeme:2012} and the recent study by \citet{Delhaize:2013}.
Two different approaches, DLA measurements and {\HI} 21-cm emission measurements, are included in the plots. Since DLA measurements normally include a correction for helium abundance, we de-corrected these values for a direct comparison of {\HI} gas density measurements. 

At $z$~$\sim$~0, two {\HI} surveys, HIPASS \citep{Zwaan:2005} and ALFALFA \citep{Martin:2010}, measured the {\HI} gas density with very good precision over large volumes of the Universe, and these measurements are in good agreement with each other. 
The preliminary result of the on-going deep {\HI} survey at redshifts up to $z$~$\sim$~0.16 with the Arecibo 305~m telescope \citep[Arecibo Ultra Deep Survey (AUDS),][]{Freudling:2011} shows {\OHI} measurement at $z \sim$~0.13 is consistent with the measurements of HIPASS and ALFALFA, taking error margins into account. \citet{Delhaize:2013} have recently measured the stacked {\HI} gas content of $\sim$~3,300 objects at $z \sim$~0.1 from a 42~deg$^{2}$ field near the South Galactic Pole (SGP) using the Parkes radio telescope. In addition, they stacked the HIPASS spectra of $\sim$~15,000 objects at $z \sim$~0.03, selected from the 2dF Galaxy Redshift Survey (2dFGRS). They also found that their {\OHI} measurements at both redshift bins agree with {\OHI} values at $z \sim$~0. 
The {\HI} gas density measurement at $z$~$\sim$~0.24 using the GMRT with the stacking technique was made by \citet{Lah:2007} with large uncertainty. At redshifts beyond $z=0.24$ the {\HI} gas density has been measured using the incidence rate of DLAs. Measurements at 0.5~$\la$~$z$~$\la$~1.5 used an indirect way to measure cosmic neutral gas density by observing an Mg\,{\sc ii}-selected sample of DLAs with the Hubble Space Telescope \citep[HST,][]{Rao:2006}. At this redshift range, the uncertainties remain very large because of the small sample size of low-$z$ DLAs \citep[e.g.][]{Meiring:2011}. 
{\HI} gas density measurements at $z$~$>$~2 come from \citet{Noterdaeme:2012}, who used a large sample of DLAs from an ongoing  survey in the spectra of quasars selected from the SDSS~DR9 \citep{Paris:2012}. Compared to \citet{Prochaska:2005,Prochaska:2009, Noterdaeme:2009}, the \citet{Noterdaeme:2012} result is higher due to a more conservative treatment of systematic effects and inclusion of DLAs with large column densities.  

As seen in Fig.~\ref{fig:omegaHI}, {\OHI} is well constrained at $z \sim $~0, and $z >$~2. Recent {\HI} 21-cm emission observations at $z \la$~0.2 show that there seems to be no evolution of {\HI} gas content in galaxies over the last 2.4~Gyr. Our observations are consistent with this observed non-evolution. During this cosmic time (the most recent 2.4~Gyr) the integral SFR of galaxies drops rapidly and one might expect that the total gas density in galaxies would  show a similar decline.  The lack of evidence for such a correlation between {\OHI}($z$) and SFR($z$) in the {\HI} 21-cm line data signifies that factors other than a proportionality of star formation with the size of the {\HI} reservoir are at play. 

In the period between look-back times of 2.4~Gyr ($z$~$\sim$~0.2) and 8~Gyr ($z$~$\sim$~1.2), the {\OHI} values are significantly higher but with large uncertainty. Knowledge of {\OHI} in this redshift range remains largely unconstrained. To disentangle the problem of large uncertainties of {\OHI} between 2.4~Gyr and 8~Gyr (0.2~$\la z \la$~1.2), more observations are required in this intermediate redshift regime. Currently, {\HI} emission line observations at $z \sim$~0.32 (3.6~Gyr) and 0.36 (4~Gyr) with the GMRT have been carried out and analysis is underway. The results from those experiments will be presented in our future papers.  
 
Theoretical models too have not yet reached a maturity level sufficient to be able to make firm predictions.  For example, \citet{Obreschkow:2009} predict that the  {\HI} mass density remains roughly constant from $z=$~0 to $z=$~1.5, however at higher redshifts ($z$~$>$~2) their predictions of {\OHI} do not reproduce the DLAs measurements. \citet{Power:2010} and \citet{Lagos:2011} also tested hydrogen gas evolution with various galaxy formation models by applying different prescriptions to assign gas masses to galaxies and showed similar results. In their models the predicted values at the redshifts where we made {\OHI} measurements ($z=$~0.1 to $z=$~0.2) are higher than our measurements but there seems to be no significant evolution from $z=$~0 to 2. The hydrodynamical simulation of \citet{Duffy:2012}, showing a mild evolution of {\HI} from $z=$~0 to $z=$~2, also does not accurately reproduce the values observed at $z=$~2. 

The discrepancy between observations and simulations points to the currently limited understanding of the evolution of the gas content of galaxies. As discussed in the introduction, the role of mechanisms such as gas accretion from the IGM and the galactic halo, infalling gas by mergers and interactions,  galactic fountains to replenish star formation, and feedback from star formation and AGN, remains to be fully understood. Measurements like the current one, as well as future surveys with the ASKAP, MeerKAT, APERTIF and ultimately the SKA should vastly improve the situation by providing robust measurements of the gas content of galaxies at intermediate redshifts and farther.

\section{DISCUSSION}

Since our measurements are based on an optically selected sample of galaxies, surface brightness selection effects may play a role. {\HI} gas-rich galaxies devoid of stars may be completely missed. But since such galaxies have not been demonstrated to exist in deep local 21-cm emission surveys \citep{Taylor:2005}, this bias is unlikely to affect our results. Future blind {\HI} surveys with forthcoming radio telescopes that can detect {\HI} emission in individual galaxies will lead to a more complete and unbiased value for {\OHI} in this redshift range.

It is very difficult to assess to what extent {\HI} self-absorption influences the {\HI} mass and mass density estimation based on {\HI} emission even with the blind {\HI} survey data \citep{Zwaan:2003,Zwaan:2005} because of a dearth of reliable measurements of intrinsic optical properties of galaxies such as inclination. According to the study of \citet{Zwaan:1997} and \citet{Braun:2012}, {\HI} self-absorption may cause an underestimation of {\OHI} by as much as 15 per~cent and 34 per~cent, respectively. If we were to apply these self-absorption corrections to our results, our main result of no evolution in  {\OHI} would remain unaffected. 

The sky coverage of our {\HI} observation is small and redshift coverages are narrow, which means that small volumes are used compared to other {\HI} surveys. These small volumes are sensitive to the effects of cosmic variance, which may bias our estimates of the cosmic {\HI} mass density ({\OHI}). However, in our calculation of {\HI} mass density, we use the optical luminosity density as a normalisation. This luminosity density is derived from the same survey that we use to select our galaxies \citep[CNOC2 survey,][]{Lin:1999}. However, the measurement is based on two widely separated CNOC2 fields, including our field CNOC2~0920+37, using much larger volumes than our WSRT {\HI} observation covered. Since our {\OHI} values are volume-normalised with the optical luminosity densities, cosmic variance due to the small sky coverage of {\HI} observations is diminished. Furthermore, the luminosity densities at $z \sim$~0.1 and 0.2 from \citet{Lin:1999} are consistent with measurements from other large volume surveys such as the SDSS and Galaxy and Mass Assembly (GAMA) \citep{Loveday:2012}. \citet{Lin:1999} also examined the variation of all parameters of luminosity function and density between the two separate fields and found that luminosity function parameters and luminosity densities from both fields are in good agreement.

\section{SUMMARY AND CONCLUSION}

We have used WSRT observations to measure the average {\HI} gas content of 155 galaxies selected from the CNOC2~0920+37 field in the  redshift ranges $z$~$\sim$~0.1 and $z$~$\sim$~0.2. Our measurement is based on signal stacking of the 21-cm emission spectra of these galaxies for which optical redshifts, multi-colour photometry, and morphological classifications are available. 
Using the optical luminosity density measured for this same sample, we can apply a volume correction to our {\HI} measurement and evaluate the cosmic {\HI} gas density, {\OHI}, at $z= $~0.1 and $z= $~0.2. 
We find that  {\OHI}~$=$~(0.33~$\pm$~0.05)~$\times$~10$^{-3}$ and {\OHI}~$=$~(0.34~$\pm$~0.09)~$\times~$10$^{-3}$, at $z$~$\sim$~0.1 and $z$~$\sim$~0.2, respectively. In addition, we make individual 21-cm emission line detections of 13 galaxies in the search volume. Our measurements are consistent with results at $z = 0$ based on large scale 21-cm surveys, as well as other measurements at $z < 0.2$. All measurements of {\OHI} at $z < 0.5$ based on 21-cm emission line observations are consistent with no evolution in the neutral hydrogen gas density over the last $\sim$4~Gyr. There is a large difference between the DLA measurements at high redshift and the 21-cm measurements at low redshift. However, it still remains to be seen whether this is an observational or a physical effect. More reliable measurements from {\HI} observations are required to constrain the evolution of {\OHI} between the local Universe and that measured through damped Lyman-$\alpha$ systems at high redshifts.

\section*{Acknowledgments}
We are grateful to an anonymous referee for helpful suggestions that improved this work. 
We thank Gyula J\'{o}zsa for his help with inspecting and fixing the WSRT archival data.  
We also thank Henk Hoekstra for his help in the planning phase of this project. The Westerbork Synthesis Radio Telescope is operated by ASTRON (Netherlands Institute for Radio Astronomy) with support from the Netherlands Foundation for Scientific Research (NWO).
Parts of this research were conducted by the Australian Research Council Centre of Excellence for All-sky Astrophysics (CAASTRO), through project number CE110001020.

\bibliographystyle{mn2e}
\bibliography{CNOC2_MNRAS_ref}

\appendix
\section{correction for incomplete sampling of the luminosity function}
Since the galaxy sample we use for 21-cm stacking is optically selected, we can only directly measure the average {\HI} content of galaxies above the sensitivity limit of the optical parent survey. Galaxies fainter than this limit, which could in principle be more gas rich, are not accounted for in our evaluation for the cosmic {\HI} mass density. We can correct for this bias by assuming that the known correlations between $\MHI/L$ and $L$ that exist in the local Universe are also valid at $z=$~0.1 and $z=$~0.2. We calculate a correction factor $f$, which is defined as the ratio of the gas richness of the total galaxy population in the subsample under study, and the measured gas richness of the galaxies above the optical detection limit:

\begin{eqnarray}
  f & = & \left( \frac{\langle \MHI \rangle }{\langle L \rangle} \right)_{\rm all} \Big / \left( \frac{\langle \MHI \rangle}{\langle L \rangle} \right)_{\rm observed} \nonumber \\
    & = & \frac{\int^\infty_0 \MHI(L)\, \phi(L)\, dL}{\int^\infty_0 L \, \phi(L) \, dL} \bigg / \frac{\int^\infty_0 \MHI(L)\, N(L)\, dL}{\int^\infty_0 L \, N(L) \, dL}, 
 \label{eq:append1}
\end{eqnarray}
where $\phi (L)$ is the Schechter luminosity function and $N(L)$ is the luminosity distribution of the observed objects.
 
We assume that the {\HI} mass-to-light ratio correlates with luminosity to the power $\beta$: 
\begin{eqnarray}
 \frac{\MHI}{L} & = & \gamma L^{\beta} \nonumber \\
{\rm or} \nonumber \\
 \MHI &= & \gamma L^{\beta+1}.
 \label{eq:append2} 
\end{eqnarray}
Then the correction factor in equation~(\ref{eq:append1}) becomes
\begin{eqnarray}
  f & = & \frac{\int^\infty_0 \gamma L^{\beta+1} \, \phi(L)\, dL}{\int^\infty_0 L \, \phi(L) \, dL} \bigg / \frac{\int^\infty_0 \gamma L^{\beta+1} \, N(L)\, dL}{\int^\infty_0 L \, N(L) \, dL} \nonumber \\
    & = & \frac{\int^\infty_0 L^{\beta+1} \, \phi(L)\, dL}{\int^\infty_0 L \, \phi(L) \, dL} \frac{\int^\infty_0 L \, N(L) \, dL}{\int^\infty_0 L^{\beta+1} \, N(L)\, dL}, 
 \label{eq:append3}
\end{eqnarray}

where we have used the Schechter function: 
\begin{equation} 
  \phi (L) dL = \phi^{*} \left(\frac{L}{L^{*}} \right)^{\alpha} \exp \left(-\frac{L}{L^{*}} \right) \frac{dL}{L^{*}}. \nonumber
\end{equation}
If we rearrange equation~(\ref{eq:append3}), the correction factor $f$ finally becomes 
\begin{eqnarray}
   f & = & \frac{\phi^{*} L^{*\alpha+1} \Gamma (2+\alpha+\beta)}{\phi^{*} L^{*} \Gamma (2+\alpha)} \frac{\int^\infty_0 L \, N(L) \, dL}{\int^\infty_0 L^{\beta+1} \, N(L)\, dL} \nonumber \\
    & = & \frac{L^{*\alpha} \Gamma (2+\alpha+\beta)}{\Gamma (2+\alpha)} \frac{\int^\infty_0 L \, N(L) \, dL}{\int^\infty_0 L^{\beta+1} \, N(L)\, dL},
\end{eqnarray}
where $\alpha$ is the faint-end slope of the optical luminosity function and $L^{*}$ is the characteristic luminosity. Both are taken from \citet{Lin:1999}. Since $\alpha$ and $L^{*}$ vary with galaxy type and redshift, a different correction factor, $f$, is calculated based on galaxy type and redshift. For the value of $\beta$ in equation~(\ref{eq:append2}), which quantifies the correlation between $\MHI/L$ and $L$, we adopt $\beta = -0.40$ from \citet{Toribio:2011}. The final calculated correction factors are listed in Table~\ref{tab:f}. A recent {\HI} survey for nearby early-type galaxies with the WSRT shows no strong correlation between $\MHI/L$ and $L$ \citep{Serra:2012}, and therefore it may be inappropriate to apply the same $\beta$ to the early-type subsample. Instead, we assume $f = 1$ for the early types, but since these galaxies make only a minor contribution to the total {\HI} gas density, the exact value of $f$ for these galaxies does not affect the conclusions.      

\begin{table}
 \caption{The correction factors $f$ as a function of galaxy type and redshift.}
 \label{tab:f}
 \centering
 \begin{tabular}{@{}lccc}
   \hline
     Sample  & $f$ ($z \sim$~0.1) & & $f$ ($z \sim$~0.2) \\
   \hline 
     Early & 0.810& & 0.995 \\
     Intermediate & 1.062 &  & 0.986 \\
     Late & 1.323 & &1.689 \\
   \hline 
 \end{tabular}
\end{table}

This correction factor derived above is very similar to the one \citet{Delhaize:2013} used for their calculation of {\OHI} based on 21-cm stacking. The important difference between our derivation and theirs is that \citet{Delhaize:2013} stack values of $\MHI/L$, while we stack {\MHI}. 

\label{lastpage}
\end{document}